\documentclass[twocolumn]{article}          
\usepackage[affil-it]{authblk}
\usepackage[utf8]{inputenc}
\usepackage{xcolor}
 \usepackage{csquotes}
 \usepackage[ngerman, english]{babel}
 \usepackage[numbers, sort&compress]{natbib}
  \usepackage{graphicx}
\usepackage[hyphens]{url}
\usepackage[hidelinks, colorlinks, linkcolor=blue, citecolor=blue, urlcolor=blue]{hyperref}
\hypersetup{urlcolor=blue}
\usepackage{wasysym}
\usepackage{amssymb}
\usepackage{upgreek}

\providecommand{\keywords}[1]
{
  \small	
  \textbf{\textit{Keywords---}} #1
}

\renewcommand{\figurename}{Fig.}
\newcommand{\figurenames}{Figs.}

\begin{document}

\title{Magnetization of Bi$_{2}$Sr$_{2}$CaCu$_{2}$O$_{8+\delta}$ micrometer thin ring and its depinning line}
 
\author[1,2]{B. Semenenko \thanks{semenenko@studserv.uni-leipzig.de}}
\author[1,3]{B. C. Camargo}
\author[1]{A. Setzer}
\author[1]{W. Böhlmann}
\author[4]{Y. Kopelevich}
\author[1]{P. D. Esquinazi \thanks{esquin@physik.uni-leipzig.de}}
\affil[1]{Felix Bloch Institute for Solid State Physics, Faculty of Physics and Earth Sciences, University of Leipzig,
Linn\'{e}stra{\ss}e 5, 04103, Leipzig, Germany}
\affil[2]{\emph{Present address:} Center of Pharmaceutical Engineering (PVZ), TU Braunschweig, Franz-Liszt-Str. 35a, 38106 Braunschweig, Germany}
\affil[3]{\emph{Present address:} Institute of Physics, Polish Academy of Sciences, Aleja Lotnikow 32/46, PL-02-668 Warsaw, Poland}
\affil[4]{Instituto de F\'{i}sica “Gleb Wataghin,” Universidade Estadual de Campinas, Unicamp 13083-970, Campinas, S\~{a}o Paulo, Brasil}

\date{Received: 8 April 2020 / Accepted: 23 May 2020}

\maketitle

\begin{abstract}
We demonstrate a geometrical effect on the depinning line (DL) of the flux line lattice of the Bi$_{2}$Sr$_{2}$CaCu$_{2}$O$_{8+\delta}$  high-T$_c$ superconductor (HTSC) micrometer ring. The DL shifts to notably lower temperatures in comparison to bulk crystals and thin flakes of the same  sample. The shift is attributed to a decrease in the overall pinning potential due to a double size effect, namely: a)  the ring thickness $\sim 1~\mu$m being smaller than the pinning correlation length, and b) the  increase in the effective London penetration depth  of the vortices (Pearl vortices). The large shift of the DL to lower temperatures may influence the suitability of this HTSC for  applications  in microstrip antennas and THz emitters.

\end{abstract}

\section{Introduction}
\label{intro}
$\mathrm{Bi_{2}Sr_{2}CaCu_{2}O_{8+\delta}}$ (Bi-2212) is a high-T$_c$ superconductor (HTSC), whose magnetic and superconducting properties are still being investigated  \cite{Huang14, Kalhor17, Kharissova14, Saito16, Sato15, Semerci16, Liao2018, Yu2019, Minami2019}.
In particular, Bi-2212 single
crystals are considered to be a key element for THz emitters devices in modern electronics \cite{Minami2019, Benseman2019, Delfanazari2019, Kleiner2019}.
Recently, numerical simulations and analytical calculations classified Bi-2212 micrometer rings as a class of 
low-loss subwavelength resonators for microstrip antennas and THz devices \cite{Kalhor17}. Although these applications are intended at zero
or low applied magnetic fields, the pinning of trapped vortices within the HTSC ring should be taken into account to minimize possible losses when electromagnetic waves of large amplitude are applied. 
 
Over the past few decades, the magnetization behaviour of superconducting rings and cylinders has been investigated theoretically \cite{Brandt97, Brandt98, Brandt98b, Navau05, Sanchez01} and experimentally  \cite{Gough93, Mrowka97,Pannetier01,SCHINDLER05,Streubel00, Xue91} both in millimeter-size HTSC rings and low-T$_{c}$ superconducting micrometer-size rings \cite{Bluhm06}.
Previous studies, done on thick Bi-2212 samples, report the presence of two thermally 
activated diffusion modes of the flux line lattice (FLL)  that determine its depinning line (DL) \cite{Ziese94}. In addition,   an anomalous mode, associated with a phase transition  at low temperatures, was observed  \cite{Kopelevich93}.
Because different sample parameters, such as the geometry and/or defect (or doping) density \cite{Gupta89_EPL, Gupta89_PRL, Esquinazi91, Gupta93,  Majer95, Kalisky2006}, influence the response of the FLL of HTSC's to external perturbations, its general description is complex.

In this work, we found a significant shift of the DL of the FLL of a micrometer-size and thin Bi-2212 ring to lower temperatures, which emphasizes the influence of the sample size (and eventually  its geometry) on the magnetic response  of this HTSC. Due to the small ring mass and consequently a low magnetic 
moment amplitude, we used a torque magnetometer to determine the DL at different applied fields and temperatures. 

\section{Experimental details}
\label{ExpDet}
 \subsection{Sample processing and structural characterization}
 \label{Sqch}
 \begin{figure}[h!]
   \centering
   \fbox{\includegraphics[width = 7.5cm]{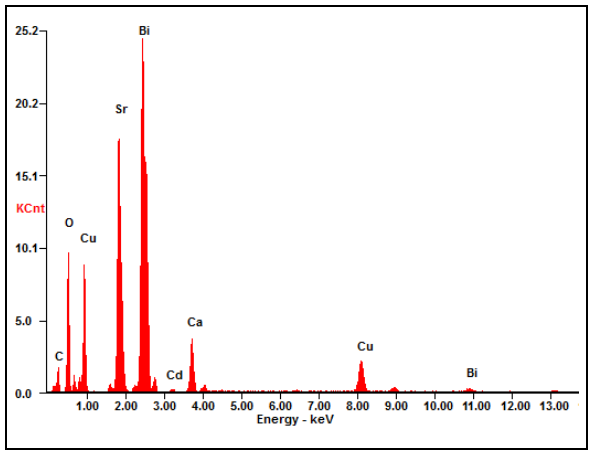}}
   \caption{EDX measurements of the  $\mathrm{Bi_{2}Sr_{2}CaCu_{2}O_{8+\delta}}$  crystal.}
   \label{fig:EDX2}
    \end{figure}

We performed energy dispersive X-ray spectroscopy (EDX) measurements to determine the composition  of 
a bulk 2212 sample, from which flakes were later extracted and processed in the shape or mesoscopic ring.
Measurements (\figurename~\ref{fig:EDX2}) were performed with an EDX detector in the Dual Beam Microscope (DBM, Nova NanoLab 200, FEI Company), using an acceleration voltage of 15 kV. The averaged values over the sample surface show a ratio of Bi/Sr/Ca/Cu elements 2/1.7/0.8/2.2 with an error of $\pm 0.2$.
The values agree with published  characterizations  on similar crystals \cite{Kopelevich1999,TORRES2003}. 
Note that the carbon peak seen in the spectrum derives from tape on which the crystal was glued and it is not an impurity of the sample.
 
 The $c$-axis of each of the measured samples is oriented in the direction normal to their main area, i.e. in the direction of their thickness.
 The thickness, volume and  mass  of all measured samples are presented in Table~\ref{table:bisko_mass}.
 The mass of the Bi-2212 large bulk crystal was measured using a Mettler Toledo AG245 balance, whereas the masses of  other samples were calculated based on the mass density 6.71 g/cm$^{3}$ \cite{Fujishiro94}.
 
\begin{table}[h!]
	\caption{The thickness, volume (V) and mass of the samples used for magnetization measurements.}
	\label{table:bisko_mass}
	\centering
	\begin{tabular}{ c   c   c   c }
		\hline\noalign{\smallskip}
		sample & thickness [$\upmu$m] & $ V$ [$\upmu$m$^3$]  & mass  [g]\\ 
		\noalign{\smallskip}\hline\noalign{\smallskip}
		bulk  & 320 & 6.5$\cdot$10$^{8}$ & 4.36$\cdot$10$^{-3}$ \\ 
		S7 & 7.44 & 9.0$\cdot$10$^5$ & 6.1$\cdot$10$^{-6}$ \\
		disk &6.7&1.3$\cdot$10$^4$&8.8$\cdot$10$^{-8}$\\
		S1 & 2.56 & 5.6$\cdot$$10^3$ & 3.7$\cdot$10$^{-8}$ \\
		ring  & 1.3 & 5.6$\cdot$10$^2$ & 3.7$\cdot$10$^{-9}$ \\
		\noalign{\smallskip}\hline
	\end{tabular}
\end{table}

\begin{figure}[h!]
	\centering
	\includegraphics[width = 7.5 cm]{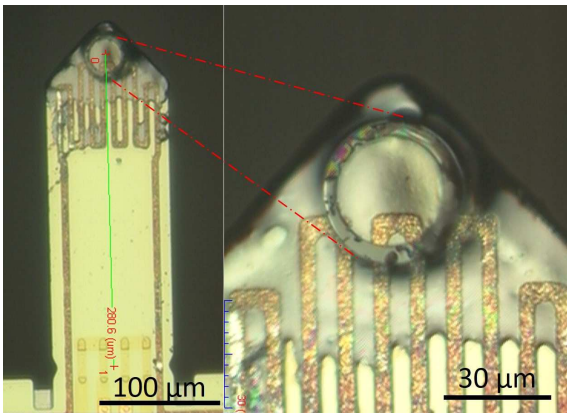}
	\caption{Optical images of the Bi-2212 micrometer ring fixed  with a cryogenic vacuum grease at the edge of the cantilever of the torque magnetometer.}
	\label{fig:ring}
\end{figure}

For the preparation of the Bi-2212 ring, shown in \figurename~\ref{fig:ring}, we deposited a Bi-2212 flake of 1.3 $\upmu$m thickness on a Si/SiN substrate and covered it with a PMMA layer of 1 $\upmu$m thickness. The protective
PMMA layer on top of the Bi-2212 flake assures that the ion beam etching procedure does not influence the material of the ring \cite{Streubel00}.
The flake with the PMMA layer was treated in air at 120$^\circ$C for the duration of 30 min.
Such treatment had a small influence on the sample critical temperature ($T_c$), as evidenced by electrical resistance measurements, shown in the inset of \figurename~\ref{fig:ln_rho_T}($a$), as well as by magnetization measurement (\figurename~\ref{fig:MvsTdisk}).
The Bi-2212 ring  was then prepared using a focused ion beam (FIB) with Ga$^+$ ions accelerated with 30~kV and with a current of  0.3~nA.  
The processing yielded a ring with a thickness $t \simeq$  1.3 $\upmu$m, 
 an outer diameter of 38 $\upmu$m and an inner diameter of 30 $\upmu$m. The width of the ring of
$w\simeq$ 4 $\upmu$m was measured with by the DBM.
The Bi-2212 disk was prepared by the same method as the ring with an outer diameter of 50 $\upmu$m and a thickness $t \simeq$  6.7 $\upmu$m.

 \subsection{Torque magnetometer}
   Because Bi-2212 HTSC has a strong magnetic anisotropy (i.e., the magnetic response along the $c$-axis $\mid$$m_{\parallel}$$\mid$ is much larger than the one in-plane $\mid$$m_{\perp}$$\mid$ \cite{Farrell89, Gu98, Haraguchi06, Musolino03, Ricketts94, Steinmeyer94}), its magnetization along the sample $c$-axis
    can be investigated with torque magnetometry.
  To measure the magnetization, we constructed a torque magnetometer with cantilevers from the company SCL-Sensor Tech. Fabrication GmbH. 
 Four piezoresistors on the cantilevers are part of  a Wheatstone bridge, with 2  piezoresistors  placed at the edge  (which measures  the cantilever deflection) and 2 piezoresistors  on the base of the cantilever (for current compensation) \cite{Keysight07}. Such configuration compensates the magnetic field influence on the piezoresistors, increasing the sensitivity of the system to magnetic moments of the order of $m$ $\sim$ 5$\cdot$10$^{-10}$~emu at 0.1~T applied field \cite{Semenenko18}. The cantilever was fixed in a holder inside
a continuous He-flow cryostat 
equipped with a superconducting solenoid powered by a IPS120-10 (Oxford) power supply. For the experiments, a resistance bridge Lake Shore 370 AC, temperature controller LakeShore 340 and Keithley 2000 multimeter were used. The angle between the magnetic field and the sample was controlled by a Hall sensor installed at the sample holder.
      
  For the torque measurements, each sample was fixed at the edge of the cantilever with vacuum grease Apiezon “H”, exemplary depicted in \figurename~\ref{fig:ring}.
  All measurements with the torque magnetometer were done with an applied magnetic field sweep rate of $\sim$ 0.14 T/min.
  Details of the calculation of the magnetic moment of the samples using this technique can be found in Appendix 1.
  
  \subsection{SQUID magnetometer}
  
  To check the calibration of the torque magnetometer, we performed measurements with a SQUID magnetometer MPMS XL-7  from Quantum Design, which allows measurements with an accuracy of $\sim$ 10$^{-8}$ emu. The sample temperature was stabilized within $\pm$ 1 mK. All measurements were done in the zero field cooled (ZFC) condition with a field sweep rate of $\sim$ 0.02 T/min.
   $T_{c}$ was obtained from M(T) measurements of the Bi-2212 bulk crystal, performed in the ZFC regime.

  \subsection{Electrical resistivity}
In order to characterize the  transport properties of the Bi-2212 material, electrical resistance measurements were performed on the sample L1 (thickness of $\simeq$ 22 $\upmu$m,  width of $\simeq$ 2.2  mm and length of $\simeq$ 0.3 mm),  taken from the same bulk Bi-2212 sample. The measurements were performed as a function of temperature at fixed magnetic fields up to 2 T with the standard 4-point method. The electrodes were made with  silver paste along the sample's $a\mbox{-}b$ plane. Measurements were performed with an AC Resistance Bridge LR-700 from Linear Research Inc., operating with a frequency of 19 Hz and a current of 10 $\upmu$A.

\section{Results}

 \subsection{Transport properties}
 
\begin{figure}[h!]
	\begin{minipage}[h]{1\linewidth}
		\hspace*{-0.6cm}
		\includegraphics[width = 7.5cm]{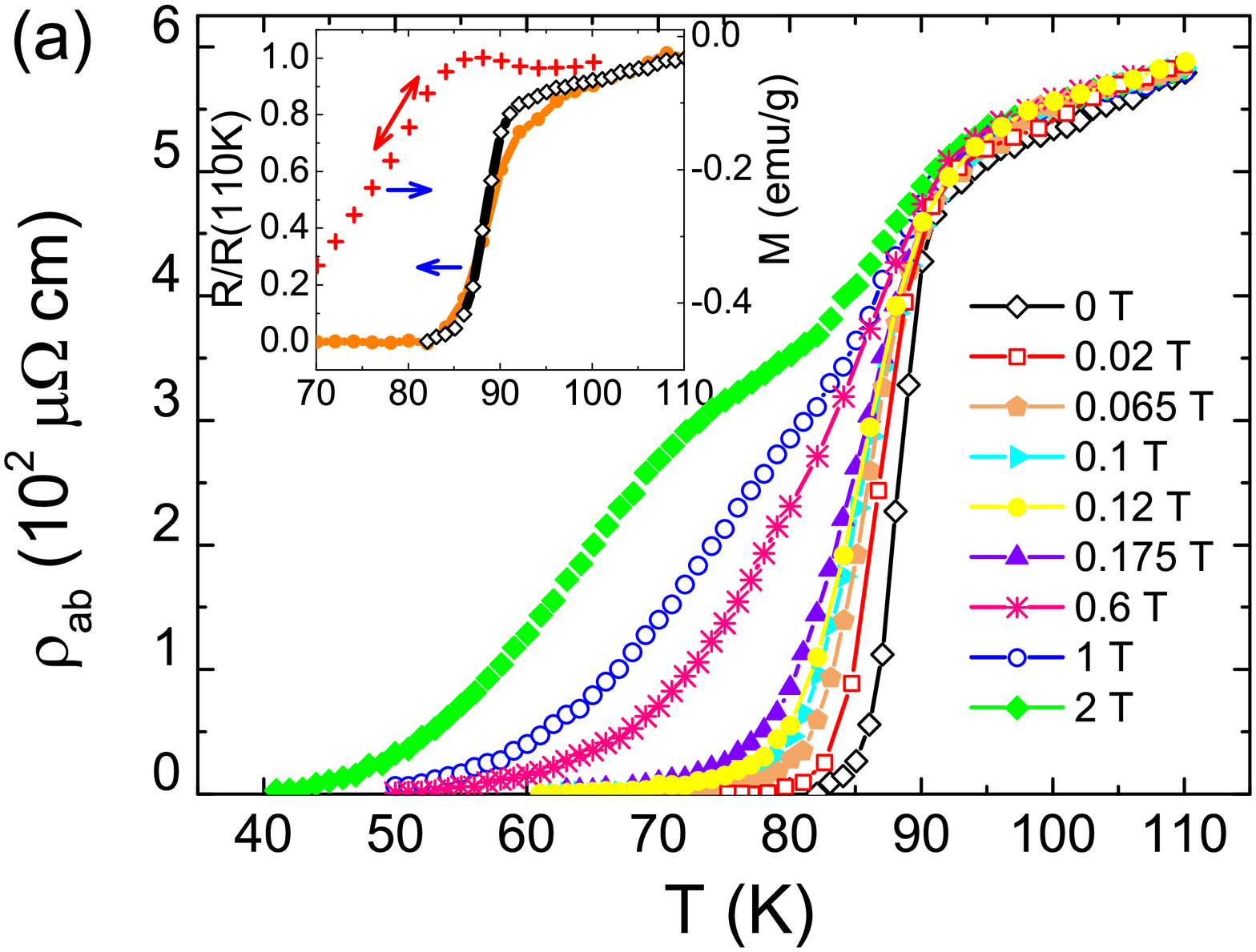} 
	\end{minipage}
	\begin{minipage}[h]{1\linewidth}
		\hspace*{-0.64cm}
		\includegraphics[width = 7.5cm]{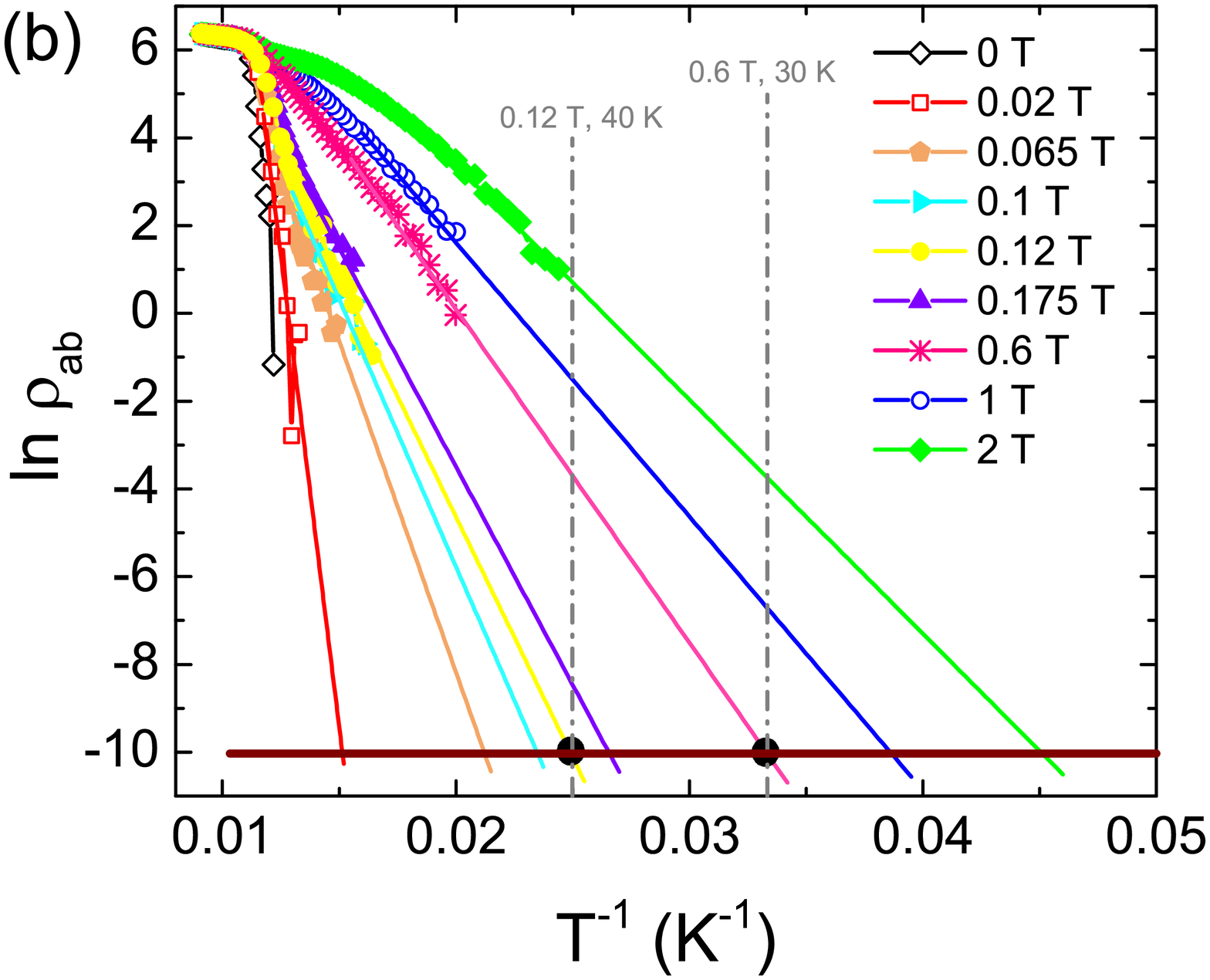} 
	\end{minipage}
	\caption{(a) Resistance vs. temperature  of the sample L1 at different  fields applied normal to the $a\mbox{-}b$ planes. 
	The inset shows the temperature dependence of the resistance before ($\textcolor{orange}{\bullet}$) and after ($\diamond$) the thermal treatment in air at 120$^\circ$C for 30 min.
   In the inset we show also the reversible part of the magnetization temperature dependence (right $y$-axis. The whole measurement can be seen in \figurename~\ref{fig:MvsTdisk} in Appendix 3) obtained at 100 Oe applied field on the disk ($\textcolor{red}{+}$). This disk was prepared following exactly the same procedure used to prepare the ring. The diamagnetic superconducting signal starts at the temperature where the resistance shows percolation, as expected.
		(b) Arrhenius plot of the same data as in (a).
		The thick horizontal line corresponds to a constant resistivity of $\rho = 4.5 \times 10^{-5}~\upmu \Upomega$cm. The intersection points between this line 
		and the extrapolated exponential behaviour provides a DL for a diffusion mode, compatible with that of the magnetization
		measurements, see \figurename~\ref{fig:depline}.}
	\label{fig:ln_rho_T}
\end{figure}

The temperature dependence of the electrical resistivity of sample L1 at different applied fields, parallel to its $c$-axis, 
is shown in \figurename~\ref{fig:ln_rho_T} (a). The critical temperature at
the middle point of the transition is $T_{c}~\simeq$~87~K at zero applied field.
The behaviour of sample resistivity against temperature at finite fields  agrees with previously reported data on the same material \cite{TORRES2003}. The high temperature resistance step clearly visible at an applied field of 2~T can be explained   within a model of weakly coupled  superconducting layers, where the motion of distorted vortices causes an excess dissipation \cite{Hsu91}. Note that this phenomenon  occurs at fields above 1~T and at high temperatures and it is not related to the thermally activated  DL, discussed here.

The mechanism known as
thermally activated flux flow (TAFF) \cite{Kes89} describes 
the thermally activated vortex motion  of the FLL 
 through a finite potential barrier $U_0$ in  the sample matrix. 
At low external or internal shielding currents, due to an external perturbation of the FLL, this mechanism leads 
to a linear (ohmic) flux flow resistivity $\rho_{TAFF}(B,T) \simeq \rho(U_0,T) exp(-U_0 /k_BT)$, with $\rho(U_0,T) \propto U_0/T$. 
In the first approximation, the main temperature dependence of $\rho_{TAFF}$ follows the Arrhenius law, i.e. 
$ln(\rho)$  
decreases linearly with $T^{-1}$, as shown in \figurename~\ref{fig:ln_rho_T}(b). This
model further predicts that a DL represents approximately a line of constant diffusivity $D = \rho_{TAFF}/\upmu_0$, resulting in a line of constant resistivity  \cite{Esquinazi91, Ziese94}.  The constant resistivity line at $\rho = 4.5 \times 10^{-5} \upmu \Upomega$cm and marked in \figurename~\ref{fig:ln_rho_T}~(b) approximately agrees with  the DL, obtained from the magnetization measurements, as  discussed below.  

\clearpage
 \subsection{Magnetization}

\begin{figure}[h!]
	\centering
	\includegraphics[width = 7.5cm]{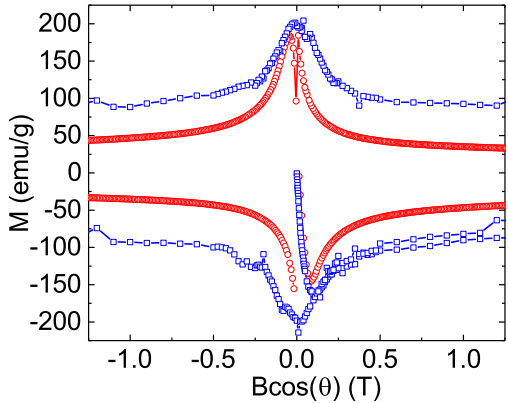}
	\caption{Magnetization hysteresis loop of the flake S1, measured with the torque magnetometer ($\textcolor{red}{\ocircle}$), and the flake S7 ($\textcolor{blue}{\square}$), measured with a SQUID magnetometer. All measurements were performed at a temperature of 10 K and at an angle of 60$^\circ$ between the $c$-axis of the samples and the applied field.}
	\label{fig:MvsB_10K}
\end{figure}

One magnetization field loop for the flake S1, measured in the torque magnetometer, is exemplary shown in \figurename~\ref{fig:MvsB_10K}. 
To verify the accuracy of such measurements and to calibrate the absolute values of  magnetic moment obtained, the flake  S7, which has similar aspect ratios, was characterized in  the SQUID magnetometer. The results for both samples are shown in \figurename~\ref{fig:MvsB_10K} and  the calibration details are discussed in Appendix 1. 

\begin{figure}[h!]
	\hspace*{-0.7cm}
	\includegraphics[width=0.58\textwidth]{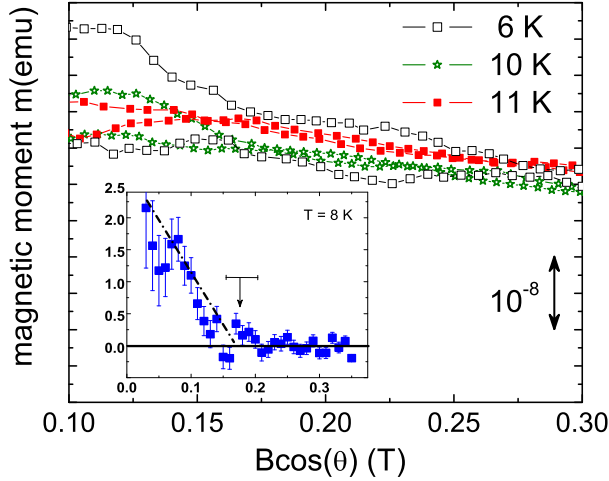}
	\caption{Magnetic moment of the Bi-2212 ring as a function of applied field at 6 K ($\square$), 10 K ($\textcolor{green}{\star}$) and 11 K ($\textcolor{red}{\blacksquare}$). 
	The  6 K data are presented here after subtraction of a diamagnetic line
	in order to show the data at the 3 temperatures in the same scale. The
	inset shows the difference between the up and down field sweep $\Delta m$ measured  on the Bi-2212 ring
	at 8 K ($\textcolor{blue}{\blacksquare}$). 
	The
	irreversibility fields were determined at the values of magnetic field where $\Delta m$ was approaching zero following \cite{Supple95}, which are $B_{irr}$ $\simeq$ 0.295 T at 6 K ($\square$), 0.175 T at 8 K ($\textcolor{blue}{\blacksquare}$), 0.17 T at 10 K ($\textcolor{green}{\star}$) with an error given by  the horizontal  bar in the middle of inset, and 0.15 T at 11 K ($\textcolor{red}{\blacksquare}$). All measurements were obtained with the torque magnetometer at the angle $\theta$ = 45$^\circ$ between the $c$-axis of the sample and  applied field.
}
	\label{fig:m_vsBring}
\end{figure}

In order to determine the DLs (or irreversibility lines) from magnetization measurements, magnetic field hysteresis loops were measured with both magnetometers at different fixed temperatures. The irreversibility field at a given temperature was  determined  at the field  where the opening of the field loop
reaches a difference between up and down field directions  $\Delta m$  of $\pm \sim 10\%$ of the largest   $\Delta m$ measured near zero field. 
Magnetization hysteresis loops, obtained for the bulk sample, the flake S1 and the disk, can be seen in \figurenames~\ref{fig:bulkMvsH},~\ref{fig:MvsHflake} and ~\ref{fig:MvsBdisk} of 
Appendix 3. The positive applied field region of the field hysteresis for the ring sample, performed at three temperatures, is shown in \figurename~\ref{fig:m_vsBring}. The depinning field was determined from the difference $\Delta m$, as exemplary shown in the inset of \figurename~\ref{fig:m_vsBring} at 8~K. Note that the  DLs of different samples were measured at different angles. Due to the significantly large anisotropy of the Bi-2212 HTSC, the DL is determined  by the field component normal to the $a\mbox{-}b$ planes of the structure, as shown in Fig.~\ref{fig:MvsB_10K_s7_0_60_deg} of Appendix 2.

In \figurename~\ref{fig:depline} we plot the DLs obtained from these
measurements for the four characterized samples, together with previously published data for bulk crystals  \cite{Ziese94}.  The DLs,
obtained for the bulk  and flake S1 samples, agree with the  DL 1 (see \figurename~\ref{fig:depline}) . 
The disk DL is slightly shifted to lower relative temperatures.
The DL obtained for the ring, however, shows a clear shift to significantly lower relative temperatures. 

\begin{figure}[h!]
 	\hspace*{-0.2cm}
 	\includegraphics[width = 7.5cm]{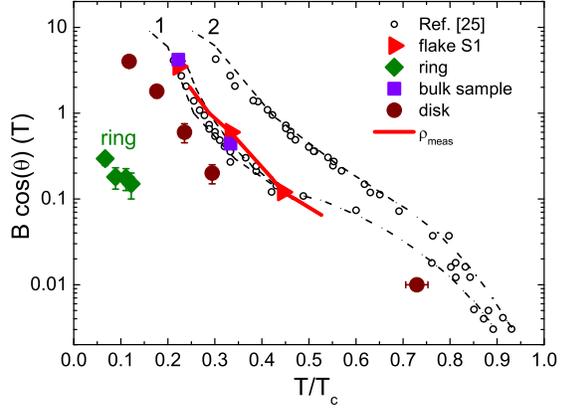}
 	\caption{DLs scaling of the Bi-2212 single crystals, shown as reduced depinning temperature ($T/T_c$) dependence of $c$-axis field component $B_{\parallel}=B\cos\theta$, where $\theta$ is the angle between the $c$-axis of the sample and the applied field  (($\ocircle$) points are taken from \cite{Ziese94}).
 		DLs scaling of the bulk crystal ($\textcolor{violet}{\blacksquare}$) and the disk ($\textcolor{brown}{\CIRCLE}$), both obtained from the magnetization hysteresis loop measurements in a SQUID magnetometer at the angle $\theta$ = 0$^\circ$;
 	 the flake S1, measured in a torque magnetometer at the angle $\theta$ = 60$^\circ$ ($\textcolor{red}{\blacktriangleright}$);
 		and the ring, measured in a torque magnetometer at the angle $\theta$ = 45$^\circ$ ($\textcolor{green}{\blacklozenge}$).
 		The solid line corresponds to the constant resistivity $\rho_{ab}$ line, obtained from resistance measurement of Bi-2212 sample L1, shown in  \figurename~\ref{fig:ln_rho_T}.}
 	\label{fig:depline}
\end{figure} 
 
 \section{Discussion}  
 
 Previous investigations \cite{Kes89,Brandt92, Gupta93, Kopelevich93, Ziese94} demonstrate that the DLs in HTSC at a given applied magnetic field (normal to the
 $a\mbox{-}b$ planes of the Bi-2212 structure) 
 can be shifted in temperature upon the corresponding diffusion time $\tau$  for a given sample geometry and measuring method, e.g.  field sweep scans, as in our case,
  or AC field (susceptibility or vibrating reed) measurements. Therefore, the diffusion time in question depends on the field sweep rate (seven times shorter in the torque measurements than those by SQUID magnetometry) and sample geometry. The DL is given by a constant diffusivity $D$ (or resistivity $\rho_{TAFF}$) line, defined as \cite{Ziese94}:
 \begin{equation}
D  \simeq l^2_i/\pi^2\tau\,,
\label{DL}
\end{equation}   
where $l_i$ is the relevant dimension of the sample where the main diffusion mode takes place. The lower DL 1 in \figurename~\ref{fig:depline} corresponds
to the diffusion mode along the thickness $t$ of the measured samples. The DLs of the flake S1 and
the bulk samples are similar within experimental error. Note that any influence of the (relatively small) difference of a factor seven  between the used sweep rates
as well as the difference in geometry are being normalized in the selected scaling of \figurename~\ref{fig:depline}. This being the reason for the universal DL obtained
in a large number of Bi-2212 samples.   
The main result of this study, therefore, is the lack of scaling and the clear shift of the DL to lower reduced temperatures obtained for the ring sample. This result  
brings additional questions to our understanding of the pinning processes in HTSC. 

The shifts of the DLs of the Bi-2212 ring and disk are not associated with the influence of the Ga$^+$ irradiation during the preparation process because of 
the following  reasons: (i) the samples were protected before preparation with a 1 $\upmu$m thick layer of PMMA, which prevents the entering of the Ga$^+$ 
ions into the main part of the samples.  (ii) In the case that some Ga doping would occur, the DL 
 shift to higher reduced temperatures might be expected
  with the increase in the number of defects  \cite{Kumakura93,  Pradhan96,  Villard96, Li97, Sun00, HARAGUCHI05, Haraguchi06, Panarina2010}. Also, the electrical resistivity measurements, as well as the magnetization measurement (inset in \figurename~\ref{fig:ln_rho_T}(a), \figurename~\ref{fig:MvsTdisk}), revealed that the annealing treatment in air nor the FIB using Ga$^+$ ions do not decrease
the $T_c$ of the sample. Furthermore, there is no decrease of $T_c$ with the reduction of the sample thickness of Bi-2212 HTSC \cite{Yu2019}. 

What about the influence of the geometrical barrier? 
A shift to lower temperatures  of the DL has been  observed at $T \geq 76$~K in Ref. \cite{Majer95}  when the Bi-2212 sample was polished from a platelet geometry into a prism shape, i.e. thereby reducing the geometrical barrier. We note, however, that the main pinning potential, that influences the DLs shown in \figurename~\ref{fig:depline}, is not related to 
the influence of the geometrical barrier, at least in the
low reduced temperature region. Moreover, we  do not expect any change in the influence of the geometrical barrier because the edge shape of the flakes or bulk samples and of the ring are similar.

It was experimentally shown that the irreversibility field can decrease in Bi-2212 single crystals and films when the thickness of the samples 
is less than the pinning correlation length $L_{0}$, which for Bi-2212 is estimated to be $\sim$ 10 $\upmu$m \cite{Matsushita05}. This phenomenon is also manifested in the DL of the disk, obtained from the magnetic measurements.
 In this case, we expect a reduction of the pinning potential $U_0$ if the thickness of the sample decreases. According 
 to Refs.~\cite{Ihara96, Matsushita2003, Matsushita05}, a
 decrease of a factor of two in the thickness $t < L_0$ between the ring and flake S1 may  produce a similar decrease in $U_0$. Moreover, the decrease of the sample thickness can produce an extra decrease in the pinning potential of the FLL if the thickness is of the order or smaller
than the London penetration depth $\lambda (\sim 4~\upmu$m at low temperatures). Under such circumstance, the penetration depth is replaced by the Pearl's effective length $\Lambda = 2\lambda^2/t \gg  \lambda$   \cite{Pearl64}. A
broadening of the effective  size of a vortex is  expected  to decrease the pinning potential and, therefore, the DL. 
These so-called Pearl vortices have been recently  visualized in narrow flat rings, made of amorphous MoGe superconducting thin films  \cite{Kokubo2019},
as well as in Y-123 HTSC thin films \cite{Acosta19}.

Therefore, the DL shift in the thin Bi-2212 ring can be interpreted as a result of the double size effect: a decrease in the pinning potential when the thickness is below the correlation length $L_0$ and the increase in the external size of the vortices (Pearl's vortices). This shift of the DL can have a detrimental effect on the possible applications of thin rings as antennas or emitters. 
A possible solution to reduce this influence could be a strong doping of the device, although the shift of the DL through doping remains rather restricted \cite{Kumakura93,  Pradhan96,  Villard96, Li97, Sun00, HARAGUCHI05, Haraguchi06, Panarina2010}.

\section{Conclusions}
In summary, we demonstrated that the DL of Bi-2212 can be significantly shifted to lower reduced temperatures. This large
detrimental effect on the DL is attributed to the double size effect, where the small thickness of the narrow ring plays a main role. 
This effect should be taken into account when considering HTSC-based devices for THz range application as proposed in Ref. \cite{Kalhor17}.

\section*{Acknowledgements}
We thank  A. Deutschinger, F. Bern and C. E. Precker for the technical assistance and M. Lorenz for helpful discussions. 
B. S. acknowledges the partial financial support from Erasmus Mundus Webb project no. 2012-27.39/001-001-EM and the partial financial support from DAAD-Programm STIBET; B. C. C. acknowledges the financial support from the National Science Center, Poland, research project
no. 2016/23/P/ST3/03514 within the POLONEZ programme. The POLONEZ programme has received funding from the European Union’s Horizon 2020 research and innovation programme under the Marie SklodowskaCurie grant agreement No. 665778. Y. K. was supported by FAPESP and CNPq.

\section*{Appendix 1: Details of torque magnetometry and the calibration of the device}
\sectionmark{Appendix A}
\label{Appendix A}
To calculate the magnetic moment of the sample from the measured resistance in the Wheatstone bridge, the following  equation was used:

\begin{equation}
m = \frac{k\Delta x l}{B\sin(\theta)},
\end{equation}
where $m$ is the magnetic moment of the sample, $\Delta x$ is the measured deflection at the edge of the cantilever tip, $\theta$ is the angle between the magnetic moment and magnetic field $B$, $l$ is the length of the cantilever tip.
Since the cantilevers have piezoresistors as a deflection sensor, their sensitivity is described by the following equation
\cite{Barlian09, KANDA91, Seto76, Yu02}:

\begin{equation} \label{eq:gauge_factor_sqrt}
\frac{\Delta R}{R_{0}}=GF\frac{\Delta x}{l},
\end{equation}
where $\Delta R$ is a resistance change of the piezoresistor, $R_{0}$ is a resistance of a piezoresistance in undeflected state, $GF$ is a gauge factor.

In order to estimate the deflection at the edge of the cantilever tip $\Delta x$, the calibration of the cantilevers were performed.
The gauge factor $GF$ of the cantilever, that was used for the Bi-2212 ring, was mechanically determined by a manipulator from SURUGA SEIKI CO., LTD, using the method described elsewhere \cite{Semenenko18, Seto76}.
Further calibration of the $GF$ was performed in the SQUID magnetometer by measuring the magnetization of the Bi-2212 flake S7, that has the identical geometry ratio \cite{Wang01} as the Bi-2212 sample S1, see \figurename~\ref{fig:MvsB_10K}. Calibration was performed along the part of hysteresis line from the zero field to the critical field $B_{c1}$, since this is the part of the loop where pinning processes do not manifest.

To determine the spring constant $k$, the equation was used from Ref. \cite{Poggi05}:

\begin{equation}
k=\frac{Et_c^{3}w}{4l^{3}},
\end{equation}
where $w$ is the width, $l$ is the length of the cantilever tip, $t_c$ is the cantilever thickness and $E$ = 1.7$\cdot$10$^{11}$ Pa is the Young modulus. 

The parameters of the cantilevers, used in our experiments, are presented in Table~\ref{table:1}.

\begin{table}[h!]
	\caption{Parameters of the cantilevers used for the measurements with the torque magnetometer: the length  ($l$), the width  ($w$), the thickness ($t_c$) and the spring constant of the cantilever tip ($k$), the resistance of the piezoresistor in undeflected state ($R_0$) and the gauge factor ($GF$).}
	\label{table:1}
	\centering
	\begin{tabular}{ p{1.3cm} p{0.4cm} p{0.4cm} p{0.45cm} p{0.6cm} p{0.6cm} p{0.4cm} p{1.1cm} }
		\hline\noalign{\smallskip}
		Cantilever No. & $l$ [$\upmu$m]& $w$ [$\upmu$m] & $t_{c}$ [$\upmu$m] & $k$ [N/m] & $R_{0}$ [k$\Upomega$] & $GF$ & sample\\ 
		\noalign{\smallskip}\hline\noalign{\smallskip}
		198  & 305 & 110 & 3.8 & 9.0 & 1.196 & 0.78 & ring\\ 
		211  & 305 & 110 & 4.0 & 10.5 & 1.165 & 0.69 &  flake S1\\
		\noalign{\smallskip}\hline
	\end{tabular}
\end{table}

We took into account that at low temperatures the geometric parameters of cantilevers do not change significantly, the Young modulus increases by $\sim$ 2 \% \cite{Greenwood88, Jeong03}, and the piezoresistive coefficients increase by $\sim$ 20 \%  \cite{Seto76}, and $R_0$ decreases by $\sim$ 20 \%.
The temperature sensitivity of the cantilevers was also taken into account to calculate  the magnetic moment of the samples.

\section*{Appendix 2: Magnetization of the Bi-2212 flake S7 at two angles}
\label{Appendix B}
\begin{figure}[h!]
	\centering
	\includegraphics[width=0.55\textwidth]{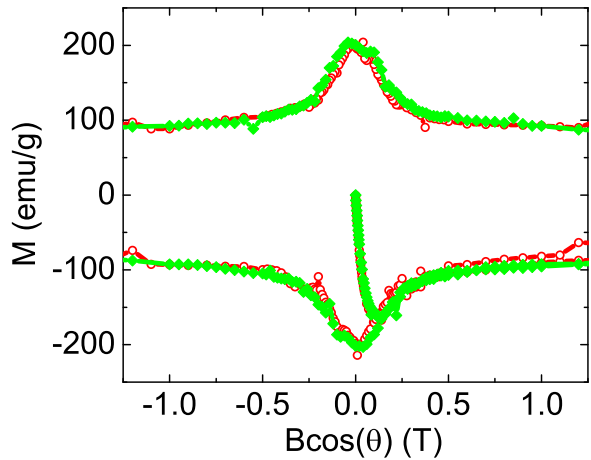}
	\caption{Field dependence of the magnetization of the Bi-2212 flake S7, measured with the SQUID magnetometer at the temperature of 10 K and at angles 0$^\circ$ ($\textcolor{green}{\blacklozenge}$) and 60$^\circ$ ($\textcolor{red}{\ocircle}$)  between the $c$-axis of the sample and the applied field.}
	\label{fig:MvsB_10K_s7_0_60_deg}
\end{figure}

Measurements of the hysteresis loops at 2 angles were carried out for the Bi-2212 flake S7 using the SQUID magnetometer (\figurename~\ref{fig:MvsB_10K_s7_0_60_deg}). These measurements indicate that the angle between the $c$-axis of sample and the field does not affect the pinning mechanism in Bi-2212 flakes.

\section*{Appendix 3: Magnetization of the Bi-2212 bulk crystal, flake S1 and disk}
\label{Appendix C}
\begin{figure}[h!]
\hspace*{-0.1cm}
	\includegraphics[width = 7.5cm]{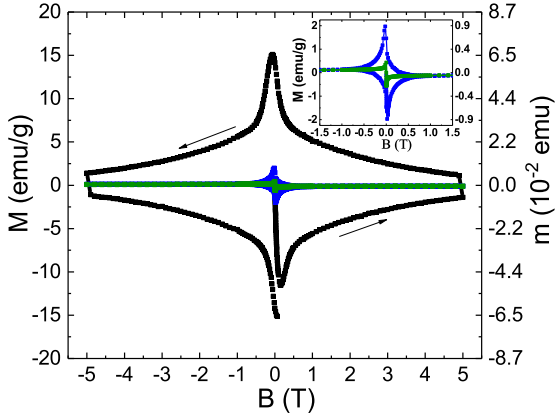}
	\caption{Magnetization of  the Bi-2212 bulk crystal versus  the applied field at $T$ = 10 K ($\blacksquare$), 20 K ($\textcolor{blue}{\CIRCLE}$) and 30 K ($\textcolor{green}{\blacktriangle}$), obtained with the SQUID magnetometer. The $c$-axis of the sample was parallel to the applied field.}
	\label{fig:bulkMvsH}
\end{figure}

\begin{figure}[h!]
\hspace*{-0.15cm}
	\includegraphics[width = 7.5cm]{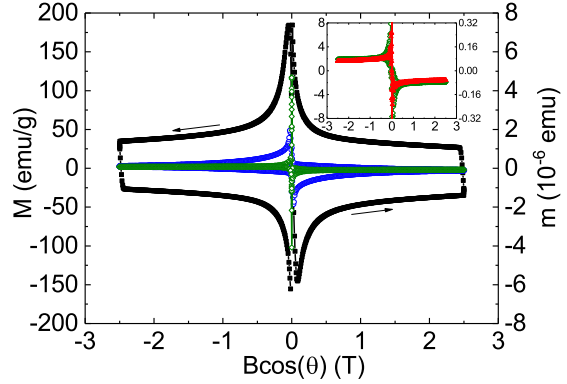}
	\caption{Magnetization of the Bi-2212 flake S1 versus the applied field at 10 K ($\blacksquare$), 20 K ($\textcolor{blue}{\ocircle}$), 30 K ($\textcolor{green}{\Diamond}$) and 40 K ($\textcolor{red}{\blacktriangle}$), obtained with the torque magnetometer at the angle $\theta$ = 60$^\circ$ between the $c$-axis of the sample and the applied field.}
	\label{fig:MvsHflake}
\end{figure}

\begin{figure}[h!]
	\hspace*{-0.15cm}
	\includegraphics[width = 7.5cm]{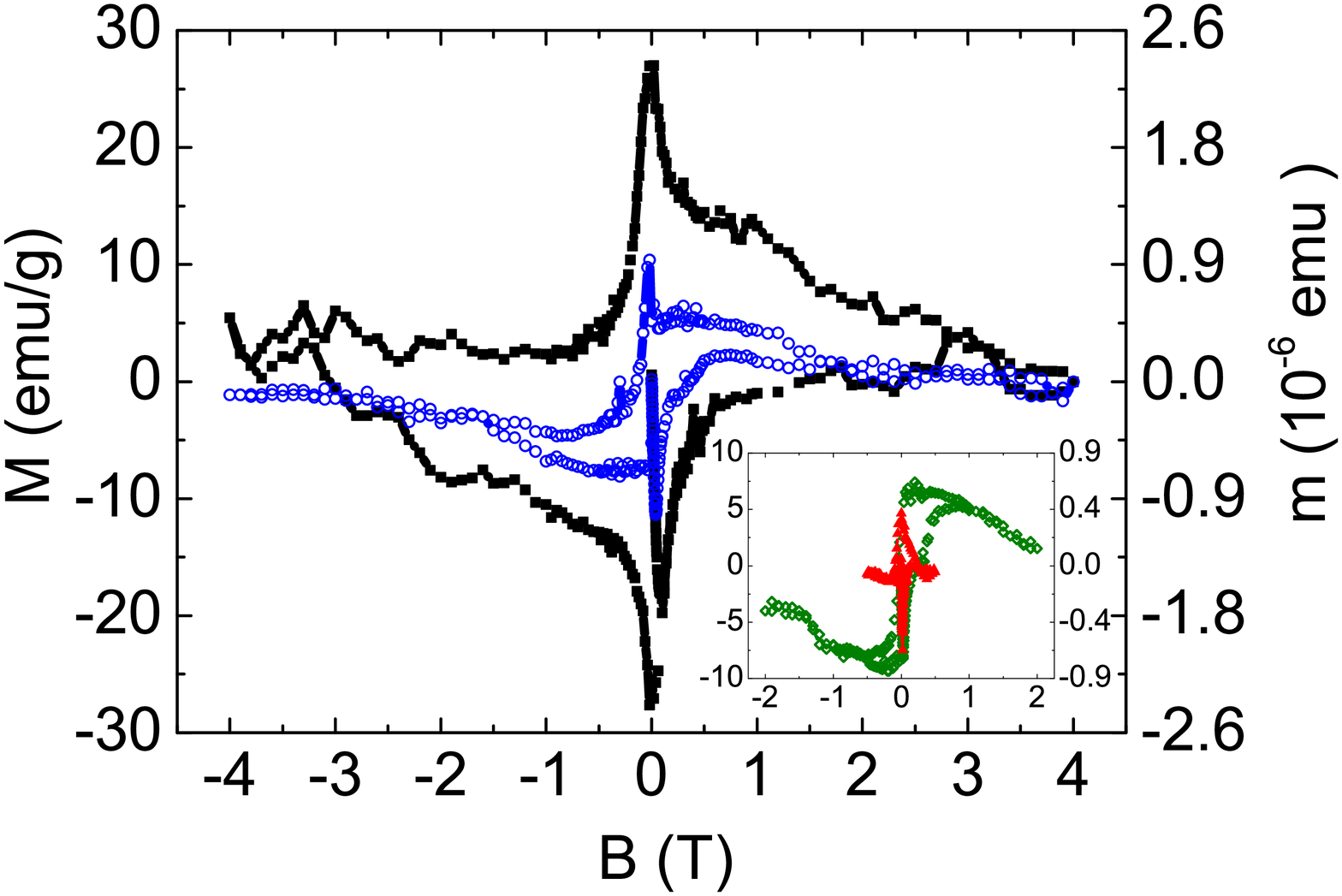}
	\caption{Magnetization of the Bi-2212 disk versus the applied field at 10 K ($\textcolor{black}{\blacksquare}$), 15 K ($\textcolor{blue}{\ocircle}$), 20 K ($\textcolor{green}{\Diamond}$) and 25 K ($\textcolor{red}{\blacktriangle}$), obtained with the SQUID magnetometer at the angle $\theta$ = 0$^\circ$ between the $c$-axis of the sample and the applied field.}
	\label{fig:MvsBdisk}
\end{figure}

The magnetization hysteresis loops for the flake S7 (\figurename~\ref{fig:MvsB_10K_s7_0_60_deg}), the bulk crystal (\figurename~\ref{fig:bulkMvsH}) and the disk (\figurename~\ref{fig:MvsBdisk}), measured by the SQUID magnetometer, and the magnetization hysteresis loops of the Bi-2212 flake S1 (\figurenames~\ref{fig:MvsB_10K} and~\ref{fig:MvsHflake}), measured by the torque magnetometer, show a typical superconducting behaviour and are comparable with the results,
reported in Ref. \cite{Dou97, FALLAHARANI18, Fossheim95, Hatta88, KISHIO91, KRITSCHA91, Noetzel96, Ries92, OZCELIK16, OZTURK18}. 

\begin{figure}[h!]
	\hspace*{-0.5cm}
	\includegraphics[width = 7.5cm]{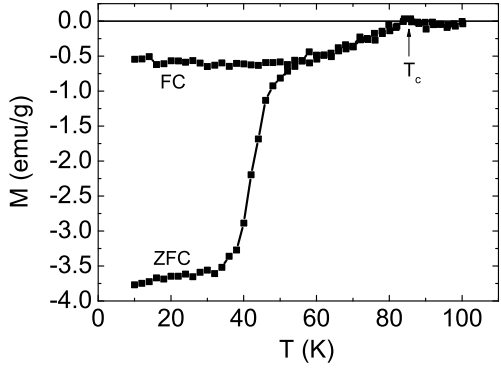}
	\caption{Magnetization of the Bi-2212 disk as a function of temperature at fixed magnetic field 100 Oe.}
	\label{fig:MvsTdisk}
\end{figure}

The ZFC and FC magnetization measurements of the disk at 100 Oe (\figurename~\ref{fig:MvsTdisk}) shows also a typical superconducting behaviour.
As can be seen from \figurename~\ref{fig:MvsTdisk}, $T_c$ of the superconducting disk is observed at $\sim$ 85 K, which indicates the absence of a significant effect of annealing at 120$^\circ$C and irradiation by gallium ions on the $T_c$ of the samples.

\clearpage

\bibliographystyle{spphys}
 \bibliography{literaturebisko,literaturecalibration}

\begin{thebibliography}{10}
\providecommand{\url}[1]{{#1}}
\providecommand{\urlprefix}{URL }
\expandafter\ifx\csname urlstyle\endcsname\relax
  \providecommand{\doi}[1]{DOI \discretionary{}{}{}#1}\else
  \providecommand{\doi}{DOI \discretionary{}{}{}\begingroup
  \urlstyle{rm}\Url}\fi

\bibitem{Huang14}
Y.~Huang, H.~Miao, S.~Hong, J.A. Parrell, IEEE Transactions on Applied
  Superconductivity \textbf{24}(3), 1 (2014).
\newblock \doi{10.1109/TASC.2013.2281063}

\bibitem{Kalhor17}
S.~Kalhor, M.~Ghanaatshoar, T.~Kashiwagi, K.~Kadowaki, M.J. Kelly,
  K.~Delfanazari, IEEE Photonics Journal \textbf{9}(5), 1 (2017).
\newblock \doi{10.1109/JPHOT.2017.2754465}

\bibitem{Kharissova14}
O.V. Kharissova, E.M. Kopnin, V.V. Maltsev, N.I. Leonyuk, L.M. León-Rossano,
  I.Y. Pinus, B.I. Kharisov, Critical Reviews in Solid State and Materials
  Sciences \textbf{39}(4), 253 (2014).
\newblock \doi{10.1080/10408436.2013.836073}

\bibitem{Saito16}
Y.~Saito, T.~Nojima, Y.~Iwasa, Nature Reviews Materials \textbf{2}(16094)
  (2016).
\newblock \doi{10.1038/natrevmats.2016.94}

\bibitem{Sato15}
K.~Sato, in \emph{Superconductors in the Power Grid}, ed. by C.~Rey, Woodhead
  Publishing Series in Energy (Woodhead Publishing, 2015), pp. 75 -- 95.
\newblock \doi{https://doi.org/10.1016/B978-1-78242-029-3.00003-0}

\bibitem{Semerci16}
T.~Semerci, Y.~Demirhan, N.~Miyakawa, H.B. Wang, L.~Ozyuzer, Optical and
  Quantum Electronics \textbf{48}(6), 340 (2016).
\newblock \doi{10.1007/s11082-016-0612-0}

\bibitem{Liao2018}
M.~Liao, Y.~Zhu, J.~Zhang, R.~Zhong, J.~Schneeloch, G.~Gu, K.~Jiang, D.~Zhang,
  X.~Ma, Q.K. Xue, Nano letters \textbf{18}(9), 5660 (2018).
\newblock \doi{doi:10.1021/acs.nanolett.8b02183}

\bibitem{Yu2019}
Y.~Yu, L.~Ma, P.~Cai, R.~Zhong, C.~Ye, J.~Shen, G.~Gu, X.H. Chen, Y.~Zhang,
  Nature pp. 156--163 (2019).
\newblock \doi{https://doi.org/10.1038/s41586-019-1718-x}

\bibitem{Minami2019}
H.~Minami, Y.~Ono, K.~Murayama, Y.~Tanabe, K.~Nakamura, S.~Kusunose,
  T.~Kashiwagi, M.~Tsujimoto, K.~Kadowaki, Journal of Physics: Conference
  Series \textbf{1293}, 012056 (2019).
\newblock \doi{10.1088/1742-6596/1293/1/012056}

\bibitem{Benseman2019}
T.M. Benseman, A.E. Koshelev, V.~Vlasko-Vlasov, Y.~Hao, U.~Welp, W.K. Kwok,
  B.~Gross, M.~Lange, D.~Koelle, R.~Kleiner, H.~Minami, M.~Tsujimoto,
  K.~Kadowaki, Phys. Rev. B \textbf{100}, 144503 (2019).
\newblock \doi{10.1103/PhysRevB.100.144503}

\bibitem{Delfanazari2019}
K.~Delfanazari, R.A. Klemm, M.~Tsujimoto, D.P. Cerkoney, T.~Yamamoto,
  T.~Kashiwagi, K.~Kadowaki, Journal of Physics: Conference Series
  \textbf{1182}, 012011 (2019).
\newblock \doi{10.1088/1742-6596/1182/1/012011}

\bibitem{Kleiner2019}
R.~Kleiner, H.~Wang, Journal of Applied Physics \textbf{126}(17), 171101
  (2019).
\newblock \doi{10.1063/1.5116660}

\bibitem{Brandt97}
E.H. Brandt, Phys. Rev. B \textbf{55}, 14513 (1997).
\newblock \doi{10.1103/PhysRevB.55.14513}

\bibitem{Brandt98}
E.H. Brandt, Phys. Rev. B \textbf{58}, 6506 (1998).
\newblock \doi{10.1103/PhysRevB.58.6506}

\bibitem{Brandt98b}
E.H. Brandt, Phys. Rev. B \textbf{58}, 6523 (1998).
\newblock \doi{10.1103/PhysRevB.58.6523}

\bibitem{Navau05}
C.~Navau, A.~Sanchez, E.~Pardo, D.X. Chen, E.~Bartolom\'e, X.~Granados,
  T.~Puig, X.~Obradors, Phys. Rev. B \textbf{71}, 214507 (2005).
\newblock \doi{10.1103/PhysRevB.71.214507}

\bibitem{Sanchez01}
A.~Sanchez, C.~Navau, Phys. Rev. B \textbf{64}, 214506 (2001).
\newblock \doi{10.1103/PhysRevB.64.214506}

\bibitem{Gough93}
C.~Gough, A.~Gencer, G.~Yang, M.~Shoustari, A.~Rae, J.~Abell, Cryogenics
  \textbf{33}(3), 339  (1993).
\newblock \doi{https://doi.org/10.1016/0011-2275(93)90056-T}.
\newblock Critical Currents in High T$_c$ Superconductors

\bibitem{Mrowka97}
F.~Mrowka, M.~Wopturlitzer, P.~Esquinazi, E.H. Brandt, M.~Lorenz, K.~Zimmer,
  Applied Physics Letters \textbf{70}(7), 898 (1997).
\newblock \doi{10.1063/1.118308}

\bibitem{Pannetier01}
M.~Pannetier, F.C. Klaassen, R.J. Wijngaarden, M.~Welling, K.~Heeck, J.M.
  Huijbregtse, B.~Dam, R.~Griessen, Phys. Rev. B \textbf{64}, 144505 (2001).
\newblock \doi{10.1103/PhysRevB.64.144505}

\bibitem{SCHINDLER05}
K.~Schindler, M.~Ziese, P.~Esquinazi, H.~Hochmuth, M.~Lorenz, K.~Zimmer,
  E.~Brandt, Physica C: Superconductivity \textbf{417}(3), 141  (2005).
\newblock \doi{https://doi.org/10.1016/j.physc.2004.10.013}

\bibitem{Streubel00}
S.~Streubel, F.~Mrowka, M.~Wopturlitzer, P.~Esquinazi, K.~Zimmer, Journal of
  Applied Physics \textbf{87}(12), 8621 (2000).
\newblock \doi{10.1063/1.373531}

\bibitem{Xue91}
Y.Y. Xue, Z.J. Huang, P.H. Hor, C.W. Chu, Phys. Rev. B \textbf{43}, 13598
  (1991).
\newblock \doi{10.1103/PhysRevB.43.13598}

\bibitem{Bluhm06}
H.~Bluhm, N.C. Koshnick, M.E. Huber, K.A. Moler, Phys. Rev. Lett. \textbf{97},
  237002 (2006).
\newblock \doi{10.1103/PhysRevLett.97.237002}

\bibitem{Ziese94}
M.~Ziese, P.~Esquinazi, H.F. Braun, Superconductor Science and Technology
  \textbf{7}(12), 869 (1994)

\bibitem{Kopelevich93}
Y.~Kopelevich, A.~Gupta, P.~Esquinazi, Phys. Rev. Lett. \textbf{70}, 666
  (1993).
\newblock \doi{10.1103/PhysRevLett.70.666}

\bibitem{Gupta89_EPL}
A.~Gupta, P.~Esquinazi, H.F. Braun, W.~Gerhäuser, H.W. Neumüller, K.~Heine,
  J.~Tenbrink, EPL (Europhysics Letters) \textbf{10}(7), 663 (1989)

\bibitem{Gupta89_PRL}
A.~Gupta, P.~Esquinazi, H.F. Braun, H.W. Neum\"uller, Phys. Rev. Lett.
  \textbf{63}, 1869 (1989).
\newblock \doi{10.1103/PhysRevLett.63.1869}

\bibitem{Esquinazi91}
P.~Esquinazi, J Low Temp Phys \textbf{85}(3-4), 139 (1991).
\newblock \doi{https://doi.org/10.1007/BF00681969}

\bibitem{Gupta93}
A.~Gupta, Y.~Kopelevich, M.~Ziese, P.~Esquinazi, P.~Fischer, F.I. Schulz, H.F.
  Braun, Phys. Rev. B \textbf{48}, 6359 (1993).
\newblock \doi{10.1103/PhysRevB.48.6359}

\bibitem{Majer95}
D.~Majer, E.~Zeldov, M.~Konczykowski, Phys. Rev. Lett. \textbf{75}, 1166
  (1995).
\newblock \doi{10.1103/PhysRevLett.75.1166}

\bibitem{Kalisky2006}
B.~Kalisky, A.~Shaulov, Y.~Yeshurun, Pramana \textbf{66}(1), 141 (2006).
\newblock \doi{10.1007/BF02704943}

\bibitem{Kopelevich1999}
Y.~Kopelevich, S.~Moehlecke, J.H.S. Torres, R.R. da~Silva, P.~Esquinazi,
  Journal of Low Temperature Physics \textbf{116}(3), 261 (1999).
\newblock \doi{10.1023/A:1021893818924}

\bibitem{TORRES2003}
J.~Torres, R.R. da~Silva, S.~Moehlecke, Y.~Kopelevich, Solid State
  Communications \textbf{125}(1), 11  (2003).
\newblock \doi{https://doi.org/10.1016/S0038-1098(02)00629-4}

\bibitem{Fujishiro94}
H.~Fujishiro, M.~Ikebe, T.~Naito, M.~Matsukawa, K.~Noto, I.~Shigaki,
  K.~Shibutani, S.~Hayashi, R.~Ogawa, Physica C: Superconductivity
  \textbf{235-240}, 1533  (1994).
\newblock \doi{https://doi.org/10.1016/0921-4534(94)91991-7}

\bibitem{Farrell89}
D.E. Farrell, S.~Bonham, J.~Foster, Y.C. Chang, P.Z. Jiang, K.G. Vandervoort,
  D.J. Lam, V.G. Kogan, Phys. Rev. Lett. \textbf{63}, 782 (1989).
\newblock \doi{10.1103/PhysRevLett.63.782}

\bibitem{Gu98}
G.D. Gu, R.~Puzniak, K.~Nakao, G.J. Russell, N.~Koshizuka, Superconductor
  Science and Technology \textbf{11}(10), 1115 (1998)

\bibitem{Haraguchi06}
T.~Haraguchi, S.~Takayama, M.~Kiuchi, E.~Otabe, T.~Matsushita, T.~Yasuda,
  S.~Okayasu, S.~Uchida, J.~Shimoyama, K.~Kishio, Physica C: Superconductivity
  and its Applications \textbf{445-448}, 123  (2006).
\newblock \doi{https://doi.org/10.1016/j.physc.2006.03.092}.
\newblock Proceedings of the 18th International Symposium on Superconductivity
  (ISS 2005)

\bibitem{Musolino03}
N.~Musolino, S.~Bals, G.~van Tendeloo, N.~Clayton, E.~Walker, R.~Flükiger,
  Physica C: Superconductivity \textbf{399}(1), 1  (2003).
\newblock \doi{https://doi.org/10.1016/S0921-4534(03)01324-8}

\bibitem{Ricketts94}
J.~Ricketts, R.~Puzniak, C.~Liu, G.D. Gu, N.~Koshizuka, H.~Yamauchi, Applied
  Physics Letters \textbf{65}(25), 3284 (1994).
\newblock \doi{10.1063/1.112438}

\bibitem{Steinmeyer94}
F.~Steinmeyer, R.~Kleiner, P.~Müller, H.~Müller, K.~Winzer, EPL (Europhysics
  Letters) \textbf{25}(6), 459 (1994)

\bibitem{Keysight07}
Keysight Technologies, \emph{Application Note: Imaging with Self-Sensing
  Cantilever on Keysight 5500/5600LS Atomic Force Microscopes.}
\newblock
  \urlprefix\url{http://literature.cdn.keysight.com/litweb/pdf/5992-2451EN.pdf?id=2902616}

\bibitem{Semenenko18}
B.~Semenenko, P.D. Esquinazi, Magnetochemistry \textbf{4}(4) (2018).
\newblock \doi{10.3390/magnetochemistry4040052}

\bibitem{Hsu91}
J.W.P. Hsu, D.B. Mitzi, A.~Kapitulnik, M.~Lee, Phys. Rev. Lett. \textbf{67},
  2095 (1991).
\newblock \doi{10.1103/PhysRevLett.67.2095}

\bibitem{Kes89}
P.H. Kes, J.~Aarts, J.~van~den Berg, C.J. van~der Beek, J.A. Mydosh,
  Superconductor Science and Technology \textbf{1}(5), 242 (1989).
\newblock \doi{10.1088/0953-2048/1/5/005}

\bibitem{Supple95}
F.~Supple, A.~Campbell, J.~Cooper, Physica C: Superconductivity
  \textbf{242}(3), 233  (1995).
\newblock \doi{https://doi.org/10.1016/0921-4534(94)02411-1}

\bibitem{Brandt92}
E.H. Brandt, Phys. Rev. Lett. \textbf{68}, 3769 (1992).
\newblock \doi{10.1103/PhysRevLett.68.3769}

\bibitem{Kumakura93}
H.~Kumakura, H.~Kitaguchi, K.~Togano, H.~Maeda, J.~Shimoyama, S.~Okayasu,
  Y.~Kazumata, Journal of Applied Physics \textbf{74}(1), 451 (1993).
\newblock \doi{10.1063/1.355279}

\bibitem{Pradhan96}
A.K. Pradhan, S.B. Roy, P.~Chaddah, D.~Kanjilal, C.~Chen, B.M. Wanklyn, Phys.
  Rev. B \textbf{53}, 2269 (1996).
\newblock \doi{10.1103/PhysRevB.53.2269}

\bibitem{Villard96}
G.~Villard, D.~Pelloquin, A.~Maignan, A.~Wahl, Applied Physics Letters
  \textbf{69}(10), 1480 (1996).
\newblock \doi{10.1063/1.116914}

\bibitem{Li97}
T.W. Li, R.J. Drost, P.H. Kes, C.~Træholt, H.W. Zandbergen, N.T. Hien, A.A.
  Menovsky, J.J.M. Franse, Physica C: Superconductivity \textbf{274}(3), 197
  (1997).
\newblock \doi{https://doi.org/10.1016/S0921-4534(96)00699-5}

\bibitem{Sun00}
Y.P. Sun, W.H. Song, B.~Zhao, J.J. Du, H.H. Wen, Z.X. Zhao, H.C. Ku, Applied
  Physics Letters \textbf{76}(25), 3795 (2000).
\newblock \doi{10.1063/1.126784}

\bibitem{HARAGUCHI05}
T.~Haraguchi, T.~Imada, M.~Kiuchi, E.~Otabe, T.~Matsushita, T.~Yasuda,
  S.~Okayasu, S.~Uchida, J.~Shimoyama, K.~Kishio, Physica C: Superconductivity
  \textbf{426-431}, 304  (2005).
\newblock \doi{https://doi.org/10.1016/j.physc.2005.01.026}.
\newblock Proceedings of the 17th International Symposium on Superconductivity
  (ISS 2004)

\bibitem{Panarina2010}
N.~Panarina, D.~Bizyaev, V.~Petukhov, Y.~Talanov, Physica C: Superconductivity
  \textbf{470}(4), 251  (2010).
\newblock \doi{https://doi.org/10.1016/j.physc.2009.09.008}

\bibitem{Matsushita05}
T.~Matsushita, M.~Kiuchi, T.~Yasuda, H.~Wada, T.~Uchiyama, I.~Iguchi,
  Superconductor Science and Technology \textbf{18}(10), 1348 (2005)

\bibitem{Ihara96}
N.~Ihara, T.~Matsushita, Physica C: Superconductivity and its Applications
  \textbf{257}(3), 223  (1996).
\newblock \doi{https://doi.org/10.1016/0921-4534(95)00534-X}

\bibitem{Matsushita2003}
T.~Matsushita, E.S. Otabe, H.~Wada, Y.~Takahama, H.~Yamauchi, Physica C:
  Superconductivity \textbf{397}(1), 38  (2003).
\newblock \doi{https://doi.org/10.1016/S0921-4534(03)01085-2}

\bibitem{Pearl64}
J.~Pearl, Applied Physics Letters \textbf{5}(4), 65 (1964).
\newblock \doi{10.1063/1.1754056}

\bibitem{Kokubo2019}
N.~Kokubo, S.~Okayasu, T.~Nojima, Journal of Applied Physics \textbf{125}(22),
  223906 (2019).
\newblock \doi{10.1063/1.5100497}

\bibitem{Acosta19}
V.M. Acosta, L.S. Bouchard, D.~Budker, R.~Folman, T.~Lenz, P.~Maletinsky,
  D.~Rohner, Y.~Schlussel, L.~Thiel, J Supercond Nov Magn \textbf{32}, 85
  (2019).
\newblock \doi{https://doi.org/10.1007/s10948-018-4877-3}

\bibitem{Barlian09}
A.A. Barlian, W.T. Park, J.R. Mallon, A.J. Rastegar, B.L. Pruitt, Proceedings
  of the IEEE \textbf{97}(3), 513 (2009).
\newblock \doi{10.1109/JPROC.2009.2013612}

\bibitem{KANDA91}
Y.~Kanda, Sensors and Actuators A: Physical \textbf{28}(2), 83  (1991).
\newblock \doi{https://doi.org/10.1016/0924-4247(91)85017-I}

\bibitem{Seto76}
J.Y.W. Seto, Journal of Applied Physics \textbf{47}(11), 4780 (1976).
\newblock \doi{10.1063/1.322515}

\bibitem{Yu02}
X.~Yu, J.~Thaysen, O.~Hansen, A.~Boisen, Journal of Applied Physics
  \textbf{92}(10), 6296 (2002).
\newblock \doi{10.1063/1.1493660}

\bibitem{Wang01}
Y.M. Wang, M.S. Fuhrer, A.~Zettl, S.~Ooi, T.~Tamegai, Phys. Rev. Lett.
  \textbf{86}, 3626 (2001).
\newblock \doi{10.1103/PhysRevLett.86.3626}

\bibitem{Poggi05}
M.A. Poggi, A.W. McFarland, J.S. Colton, L.A. Bottomley, Analytical Chemistry
  \textbf{77}(4), 1192 (2005).
\newblock \doi{10.1021/ac048828h}.
\newblock PMID: 15859006

\bibitem{Greenwood88}
J.C. Greenwood, Journal of Physics E: Scientific Instruments \textbf{21}(12),
  1114 (1988).
\newblock \urlprefix\url{http://stacks.iop.org/0022-3735/21/i=12/a=001}

\bibitem{Jeong03}
J.~hyun Jeong, S.~hoon Chung, S.H. Lee, D.~Kwon, Journal of
  Microelectromechanical Systems \textbf{12}(4), 524 (2003).
\newblock \doi{10.1109/JMEMS.2003.811733}

\bibitem{Dou97}
S.X. Dou, X.L. Wang, Y.C. Guo, Q.Y. Hu, P.~Mikheenko, J.~Horvat, M.~Ionescu,
  H.K. Liu, Superconductor Science and Technology \textbf{10}(7A), A52 (1997)

\bibitem{FALLAHARANI18}
H.~Fallah-Arani, S.~Baghshahi, A.~Sedghi, D.~Stornaiuolo, F.~Tafuri,
  N.~Riahi-Noori, Physica C: Superconductivity and its Applications
  \textbf{548}, 31  (2018).
\newblock \doi{https://doi.org/10.1016/j.physc.2018.01.012}

\bibitem{Fossheim95}
K.~Fossheim, E.D. Tuset, T.W. Ebbesen, M.M.J. Treacy, J.~Schwartz, Physica C:
  Superconductivity \textbf{248}(3), 195  (1995).
\newblock \doi{https://doi.org/10.1016/0921-4534(95)00382-7}

\bibitem{Hatta88}
S.~ichiro Hatta, K.~Hirochi, H.~Adachi, T.~Kamada, Y.~Ichikawa, K.~Setsune,
  K.~Wasa, Japanese Journal of Applied Physics \textbf{27}(9R), 1646 (1988)

\bibitem{KISHIO91}
K.~Kishio, S.~Komiya, N.~Motohira, K.~Kitazawa, K.~Yamafuji, Physica C:
  Superconductivity \textbf{185-189}, 2377  (1991).
\newblock \doi{https://doi.org/10.1016/0921-4534(91)91313-S}

\bibitem{KRITSCHA91}
W.~Kritscha, F.~Sauerzopf, H.~Weber, G.~Crabtree, Y.~Chang, P.~Jiang, Physica
  C: Superconductivity \textbf{179}(1), 59  (1991).
\newblock \doi{https://doi.org/10.1016/0921-4534(91)90011-M}

\bibitem{Noetzel96}
R.~Noetzel, B.~vom Hedt, K.~Westerholt, Physica C: Superconductivity
  \textbf{260}(3), 290  (1996).
\newblock \doi{https://doi.org/10.1016/0921-4534(96)00146-3}

\bibitem{Ries92}
G.~Ries, H.W. Neumuller, W.~Schmidt, Superconductor Science and Technology
  \textbf{5}(1S), S81 (1992)

\bibitem{OZCELIK16}
B.~Özçelik, O.~Nane, A.~Sotelo, M.~Madre, Ceramics International
  \textbf{42}(2, Part B), 3418  (2016).
\newblock \doi{https://doi.org/10.1016/j.ceramint.2015.10.137}

\bibitem{OZTURK18}
H.~Öztürk, S.~Safran, Journal of Alloys and Compounds \textbf{731}, 831
  (2018).
\newblock \doi{https://doi.org/10.1016/j.jallcom.2017.10.095}

\end{thebibliography}
\end{document}